# Usability Dimensions and Behavioral Intention to Use Markdown-to-Moodle in Test Construction

**Julius G. Garcia**
*Technological University of the Philippines, Philippines*
julius.tim.garcia@gmail.com

**Connie C. Aunario**
*Technological University of the Philippines, Philippines*
connniecruzaunario@yahoo.com

**Go Frendi Gunawan**
*STIKI Malang, Indonesia*
gofrendi@stiki.ac.id

*Creating test with numerous items in Moodle can be tedious and less intuitive compared to conventional method. This study aims to determine the Markdown-to-Moodle performance in easing the test construction process and explain the underlying factors of the behavioral intention to use the application. Markdown-to-Moodle is an application that allows users to type the bulk of test items directly to the browser and generates \*.doc, \*.md and \*.xml files stored in the local drive. The \*.xml can be imported to Moodle test bank. This lessens the time of creating test items one at a time in the Moodle. A training and a survey were conducted among teachers with Moodle usage experience. Results from this study allowed the researchers to determine the usability of the application and the user's behavioral intention. This highlights the workflow continuity in test construction as a key factor in the usage and performance of the application.*

*Keywords: Test Construction, Markdown, Moodle, Behavioral Intention, Usability Dimensions*

## Introduction

Moodle is one of the popular and well-known open source learning management system (Fenton, 2018) in the academic sector. Moodle can be implemented in a local premise for free and is available on cloud, but both have limitations. This free service is limited to 50 users and restricts the users to certain extent of its functionality and features. It also entails costs for any changes in the free system workflow ("Moodle pricing comparison table," 2019).

Currently, Moodle has 158,107,687 users and 108,385 active sites registered from 229 countries. Moreover, there are 18,701,289 courses available, 727,776,333 enrolments, 332,106,792 forum posts, 165,251,359 resources and 1,599,246,553 quiz questions ("Moodle Statistics," 2019). The data reveal that Moodle is highly utilized in test construction.

However, Moodle poses some issues when writing and constructing test items in the system. Aside from the problem that the test construction confronts such as "what to measure?" and "how to measure?" (Lindquist, 1936), the processes within these questions are activities specifically encoding and generating test items have been given less attention. Writing bulk questions in the Moodle system could be tedious for users due to the number of questions, including its choices, that needs to be encoded one at a time.

In connection to this challenge, Markdown-to-Moodle application was developed to address the tedious process of test construction in the Moodle and provide users a platform where they can directly create bulk of test questions and convert it to *.doc, *.md and *.xml files.

## Review of Related Literature

### Usability Dimensions

The usability models provide a conceptual view of the criteria or focus area to establish the usability of a software. ISO standard (ISO 9241-11, 1998) defines usability as "the degree to which a product can be used by specified users to achieve specified goals with effectiveness, efficiency and satisfaction in a specified context of use." This has been extended to five dimensions by Quesenbery (2001) namely efficient, engaging, error tolerant and easy to learn.

**Effectiveness**

Effectiveness refers to the system completeness and accuracy which helps users achieve their goals. The system performance measures to accomplish a specified task within the required time (ISO 9241-11, 1998). Effectiveness also describes how users accomplish the specified task with minimum effort.

It also investigates how the user's goals are meet using the software. It describes the systems performance at a required level and the required percentage of the user's target range within required portion of the usage environment range (Shackel, 1991).

**Efficiency**

Efficiency refers to how fast the goals are meet with accuracy and completeness by the users using the system. Efficiency refers to how directly and quickly those goals can be achieved with accuracy with which users can complete their tasks (Quesenbery, 2001). Efficiency refers to the total resources used to complete a task effectively. These resources are the user's number of individual actions taken, including the time spent on them. Nielsen (1993) directly associate efficiency to productivity. The more efficient the system, is the higher the throughput.

However, the boundaries of the task should be clearly defined when measuring efficiency. The user's perception of the complete task should be used rather than individual functions. Mainly, this issue is critical when a task involves multiple functions or when the entire task cannot be completed within the product (Quesenbery, 2001).

**Engagement**

Engagement is the degree of the system user interface style and tone that makes it pleasant, satisfying or appealing to use. This means that the systems can draw the user and engage them to create interaction. Engagement is the most subjective of the five dimensions (Quesenbery, 2001). Although the aesthetic elements of a system can generally attract its audience, other elements such as the media used, the language choice, and the interaction style also play a part in creating the continuous experience that leads to engagement. Engagement was used instead of satisfaction to highlight the user's sense of dynamic interaction and emotional level (Quesenbery, 2001).

**Error Tolerance**

Error tolerance refers to the system's ability to prevent errors and to recover from it. Nielsen (1993) states that the error rate in the system should be less and in case an error occurred, the system should be able to recover from it. An error-tolerant system helps the user recover from errors by providing information or course of actions on how to correct it (Quesenbery, 2001).

**Ease of Learning**

Ease of Learning refers to how well the system supports both initial orientation and deepening understanding of its capabilities (Quesenbery, 2001). Ease of Learning is the effort required to understand and operate an unfamiliar system (Eason, 1984).

However, Nielsen (1993) used "learnability" rather than ease of learning which means that the system should be easy to learn and understand. Likewise, learnability is the software capability to enable the user to learn its application with effectiveness, efficiency, freedom from risk and satisfaction in a specified context of use (ISO/IEC 25010, 2011).

## Behavioral Intention to Use

Behavioral intention refers to the perceived possibility of a person to engage in a behavior (Oliver, 1997). It is also defined as "the strength of one's intention to perform a specified behavior" (Fishbein & Ajzen, 1975, p. 288). This is shaped by the user's satisfactory and pleasurable experience of a product or service (Ali, Omar, & Amin, 2013; H. Kim, Park, M. Kim, & Ryu, 2013). Moreover, satisfaction plays a significant role as a predecessor for positive behavioral intention (Oliver,1997) and causes favorable intentions to use or acquire the product again or revisit the service location (Han, Hsu, Lee, & Sheu, 2011).

Behavioral intention has a positive effect on behavior that has been reflected in research studies on Technology Acceptance Model (TAM) (Lu, Lin, & Chen, 2017). The learner behavioral intention to use a system reflects system acceptance (Lee & Lehto, 2013). The various constructs influencing behavioral intention to use a system is an intent indicator of the user's understanding in the information systems analysis, design and implementation (Guo, Goh, Luyt, Sin, & Ang, 2015; Pituch & Lee, 2006; Saadé, Tan, & Nebebe, 2008). Behavioral intention is a significant predictor of action (Hill, Smith, and Mann,1987). Behavioral intention to use (BIU) is a significant construct that determines and ascertains whether the application will be utilized or not.

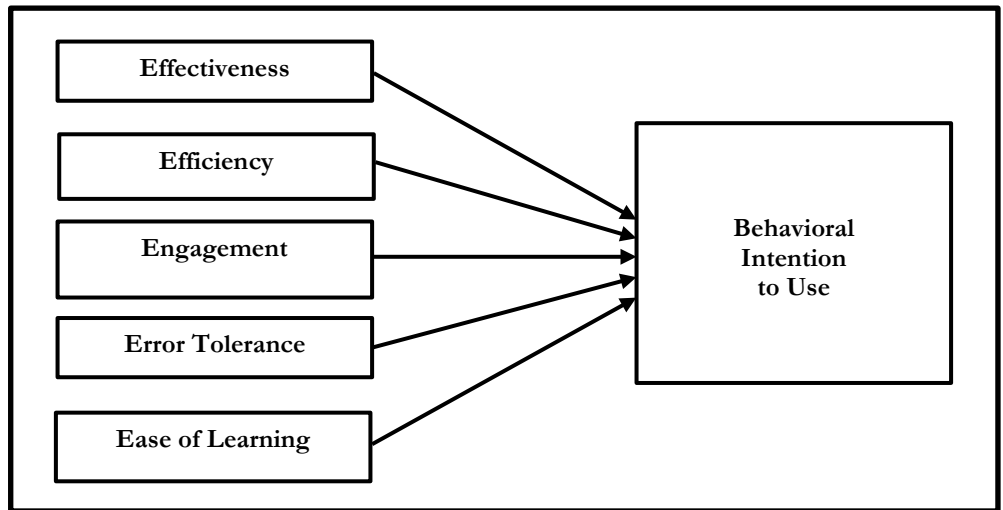

*Figure 1*. Research Framework of the Study

In this context, the usability dimensions could determine the potential of the application in item test construction. Moreover, effectiveness, efficiency, engagement, error tolerant and ease of learning could be adopted as key predictors in identifying the behavioral intention to use the application as shown in Figure 1.

**Problem Statement**

In this light, this study was conducted to determine the performance of Markdown-to-Moodle application in easing the test construction process and explain the underlying factors of the behavioral intention to use the application.

Specifically, the study sought to answer the following questions.
1. What are the internet experience and proficiency of teachers beneficial to the use of the application?
2. Is there a significant improvement on the time constructing the test items using the Markdown-to-Moodle Application?
3. What is the user's assessment on the Markdown-to-Moodle application in terms of usability dimensions; effectiveness, efficiency, engagement, error tolerance, and ease of learning?
4. What are the factors that significantly affect the behavioral intention to use the Markdown-to-Moodle application?

# Research Design And Method

**Participants and Procedure**

Training and survey were conducted respectively to private school teachers from April to June of 2018. The participants have prior knowledge and experience in using Moodle. A snowball sampling was used because the participants were invited through a referral scheme.

The training was conducted to orient the participants with the Markdown-to-Moodle application and to provide them the knowledge on how to utilize it. The Markdown format, rules in creating text, formulas and image link questions were discussed and demonstrated to the participants.

There were two sets of test items or questions (n=10) that has been given to the teachers to construct. The questions were item-response multiple choice type. Set 1 and Set 2 are 10-item questionnaires which have a similar test item format.

In set 1, the facilitator or administrator supervised the participants in constructing the test items, converting it to XML file and up until importing it to the Moodle test bank. The teachers directly created each test item using the Markdown format in the workspace of the Markdown-to-Moodle application.

In set 2, the participants were not given assistance and supervision in constructing the test items up until importing the converted XML files to the Moodle test bank and running the test questionnaire in Moodle. The training activities and tasks were recorded and collected.

A survey was conducted to the participants after the training. There were 120 valid responses. The demographic data of the participants were also collected as shown in Table 1.

Table 1

*Participants Characteristics*

| Participants Characteristics | | Results |
|---|---|---|
| Age | Mean | 31.66 |
| | S.D. | 7.704 |
| Gender | Male | 40.80% |
| | Female | 59.20% |
| Nationality | Filipinos | 86.70% |
| | Indonesians | 11.70% |
| | Cambodians | 1.70% |
| Highest Educational Attainment | Bachelors | 39% |
| | Masters | 50% |
| | Doctorate | 10.80% |
| Teaching Academic Level | Elementary | 25% |
| | Secondary | 25% |
| | Tertiary | 50% |

**Measurement of Variable**

The survey was conducted right after the facilitation and training on the Markdown-to-Moodle application. The measurement of the usability dimensions of the application was adopted from (Quesenbery, 2001) and the measurement of the behavioral intention was adopted from Ajzen (1991). The questionnaire was composed of 1) Demographics, (k = 5); 2) Effectiveness (k = 4); 3) Efficiency (k = 2); 4) Level of Engagement (k = 2); 5) Error Tolerance (k = 3); 6) Ease of Learning (k = 3); and 7) Behavioral Intention to Use the System (k = 2). We adopted a validated scale to develop our survey questionnaire, employing a seven-point scale (Vagias, 2006), with 7 as the highest and 1 as the lowest extent of agreement or frequency of use.

Descriptive statistics were used to analyze and represent the data. Learning gain was computed to measure the participant's growth or learning from the intervention period (Colt, Davoudi, Murgu & Rohani, 2011; Steif & Dollár, 2009). A t-test was conducted to determine the significant improvement in test construction. A Durbin-Watson test was also performed to determine the correlations between errors. An ANOVA test was also used to identify whether the model is significantly better at predicting the outcome. A multiple regression analysis was employed using stepwise method to examine the relationship of the predictors to the criterion.

**Markdown-to-Moodle Application Utilized**

Markdown-to-Moodle Application is a web-based application that allows teachers to directly encode the set of test questions in the web browser and generates a word, pdf and xml file. The application can be utilized with the provision of the internet. Figure 2 presents the application's interface.

**Markdown-to-Moodle Application Workflow**

The converted file is processed in md_script_to_dictionary procedure. After the conversion has been done, the weight of every option in a question will also be automatically calculated. For example, if a question has two correct answers, the weight of every correct option is 100/2. Because of Moodle rule, the weight should be written with 7 digits behind the dot. This process is done inside completing_dictionary procedure.
After the dictionary has been completed, the program will generate XML representation for every test session defined. This is done inside section_to_xml procedure. The process from the local computer to Moodle site is shown in Figure 3.

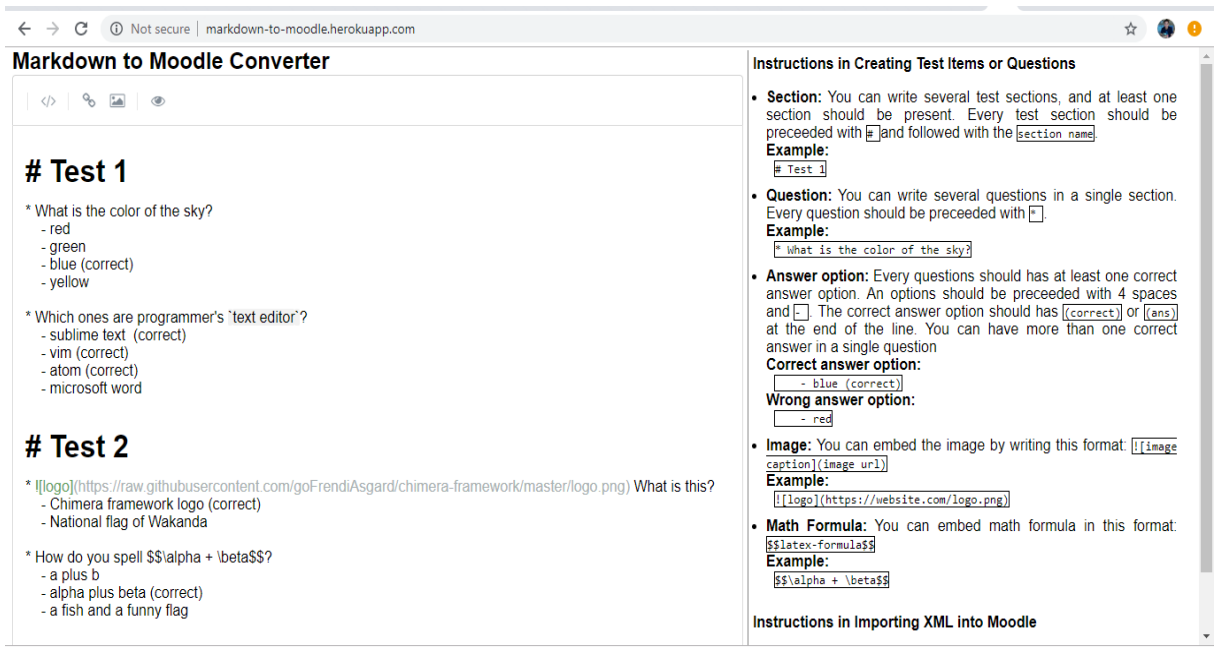

*Figure 2.* Markdown-to-Moodle Interface

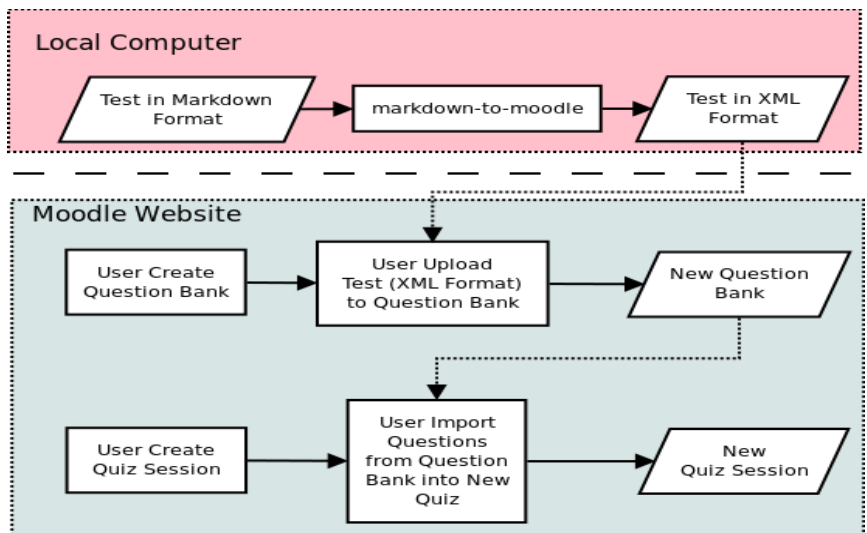

*Figure 3.* Markdown to Moodle Online Test Item Construction Flow

For every test session, an XML file is written to the local drive which could then be imported to the Moodle system. Importing the converted files into the Moodle system is another process as shown in Figure 4.

In the import questions page, select the XML format and upload your XML file. Once it is uploaded, the system will display each test items in the Moodle Test Bank.

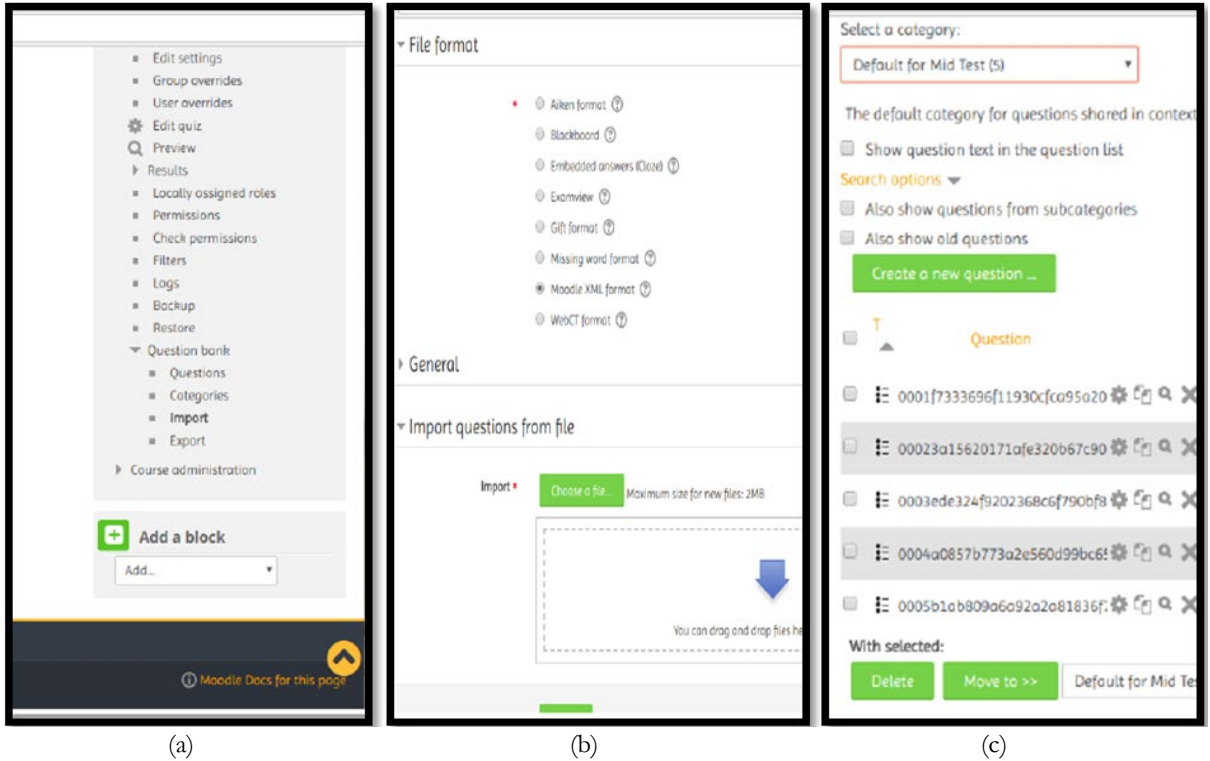

*Figure 4.* Importing XML File to Moodle Data Bank (a) Import Test Items (b) Select XML Format (c) Test Bank Item List

# Results

This study obtained 120 valid responses from private school teachers from the elementary, secondary and tertiary level based on the following: Filipino = 104 (Male = 42, Female = 62, mean of age = 32.02), Indonesian = 14 (Male = 7, Female = 7, mean of age = 29.07) and Cambodian = 2 (Male = 0, Female = 2, mean of age = 30.5); 48 male and 72 female respondents.

## Internet Proficiency and Experience of Teachers

The teachers are generally proficient (M=6.06, SD=.86) in using the internet with elementary teachers (M=5.77, SD=.77) and secondary teachers (M=6.10, SD=.75) both very proficient and tertiary teachers (M=6.18, SD=.93) extremely proficient as shown in Table 2. Moreover, they have an average of fourteen years (M=14.24, SD=1.50) in using the internet.

Table 2

*Teachers' Internet Proficiency*

| Academic Level | Mean | SD | Verbal Interpretation |
|---|---|---|---|
| Elementary(n=30) | 5.77 | .774 | Very Proficient |
| Secondary(n=30) | 6.10 | .759 | Very Proficient |
| Tertiary(n=60) | 6.18 | .930 | Extremely Proficient |
| Total | 6.06 | .863 | Very Proficient |

*Legend.* 6.16-7.00 = Extremely Proficient, 5.30-6.16 = Very Proficient, 4.44-5.29 = Moderately Proficient, 3.58-4.43 = Neutral, 2.72-3.57 = Slightly Proficient, 1.86-2.71 = Low Proficient, 1.00-1.85 = Not Proficient at All

## Experiment Results

The interaction of the teachers with the Markdown-to-Moodle Application was recorded and the time stamps were collected. The average time stamp of set 1 (M=22.86, SD=5.20) and set 2 (M=15.15, SD=4.18) as shown in Table 3. There is a decrease in time (M=7.71, SD=4.61) between set 1 and set 2 which is associated to the teachers learning gain (M=0.24, SD=0.33) of 24%. The teachers can complete constructing and converting the test items up until importing it to the Moodle test bank in lesser time. It also shows that the teachers adapt well to the rules in constructing the test items. This is also evident in the t-test result (M=6.30, SD=6.30), t(120)=15.42, p<.001 as shown in Table 4.

Table 3

*Time Stamp Descriptive Statistics on Constructing and Converting Test Items in Markdown-to-Moodle Application*

| Time Stamp: Test Items Construction to File Conversion | Mean | SD |
|---|---|---|
| Set 1 Test Items (n=10) | 22.86 | 5.20 |
| Set 2 Test Items (n=10) | 15.15 | 4.18 |
| Set 1 and Set 2 Time stamp Difference | 7.71 | 4.61 |
| Timestamp Learning Gain | 0.24 | 0.33 |

Table 4

*Time Stamp T-Test Result on Constructing and Converting Test Items in Markdown-to-Moodle Application*

| Time Stamp: Test Items Construction to File Conversion | Mean | SD | t value | Sig |
|---|---|---|---|---|
| Set 1 and Set 2 Timestamp Difference | 6.30 | 4.47 | 15.42 | .000 |

*Note.* *$p<.05$, **$p<.01$, ***$p<.001$

After accomplishing set 1, the teachers had minor errors and were able to recover with their mistakes due to the applications ability to store previous work and the guidelines indicated in the interface. It was also easier for them to learn the interface because of the application's clarity and simplicity of tasks to operate.

**Assessment Results**

The six items had an acceptable Cronbach Alpha α = 0.75 based on George and Mallery (2003) guidelines, indicating that they are correlated.

Mean scores in terms of effectiveness (M=6.26, SD=0.47) and level of engagement (M=6.24, SD=0.54) are high as shown in Table 5. This means that the teachers strongly agree that tasks given to them to construct and convert the test items in the Markdown-to-Moodle application and import the test items to Moodle were fully completed and have met the expected results.

Moreover, teachers had a pleasant experience in using the web application and were satisfied with how the web application supported their task in constructing the test items. The teachers agree that the Markdown-to-Moodle Web Application is efficient (M=5.96, SD=.68), error tolerant (M=5.93, SD=.51) and easy to learn (M=5.91, SD=.56).

Table 5

*Descriptive Statistics of Behavioral Intention to Use and Usability Dimension*

| Constructs | Mean | SD | Verbal Interpretation |
|---|---|---|---|
| Behavioral Intention to Use | 6.03 | .68 | Agree |
| Effectiveness | 6.26 | .47 | Strongly Agree |
| Efficiency | 5.96 | .68 | Agree |
| Level of Engagement | 6.24 | .44 | Strongly Agree |
| Error Tolerance | 5.93 | .51 | Agree |
| Ease of Learn | 5.91 | .56 | Agree |

*Legend.* 6.16-7.00 = Strongly Agree, 5.30-6.16 = Agree, 4.44-5.29 = Somewhat Agree, 3.58-4.43 = Neither Agree or Disagree, 2.72-3.57 = Somewhat disagree, 1.86-2.71 = Disagree, 1.00-1.85 = Strongly Disagree

Multiple linear regression was conducted using the stepwise method. It generated two models, as shown on Table 6. The first model has a value of R2 =.137, which means the error tolerance accounts for 13.7% of the variation in the behavioral intention of the use of Markdown-to-Moodle Application. On the hand, the second model increases to 18.2% or value of R2=.182, a variance which can be attributed to the behavioral intention of using the Markdown-to-Moodle Application. In addition, the Dublin-Watson value is 1.903 which indicates a positive correlation between adjacent residuals.

An ANOVA test determined that the initial model (F(1,118)=18.77, p<.001) and the second model ( F(2,117)=12.99, p<.001) are both significant. Thus, the items or predictors in the model have significant influence on the behavioral intention of using the application and that the assumptions were met.

Table 6

*The behavioral intention to use the Markdown-to-Moodle Application on the Usability Dimensions.*

| Predictors | b | SE b | β |
|---|---|---|---|
| Step 1 | | | |
| Constant | 3.124 | .672 | |
| Error Tolerance | .490 | .113 | .370*** |
| Step 2 | | | |
| Constant | 2.389 | .719 | |
| Error Tolerance | .387 | .118 | .293** |
| Efficiency | .225 | .089 | .225* |

*Notes*. $R^2$=.14 for Step 1: $\Delta R^2$=.05 for Step 2 *(ps<0.05) \*p<.05, \*\*p<.01, \*\*\*p<.001*

For step 1, the Error Tolerance, β = .370, *t*(120)=3.28, *p*<0.001 is the predictor that is making the most significant contribution to the behavioral intention of using the Markdown-to-Moodle Application. Moreover, step 2 still includes predictor Error Tolerance, β = .293, *t*(120)=2.53, *p*<0.01 and adds Efficiency, β = .225, *t*(120)=2.53, *p*<0.05 in the predictors that contribute to the behavioral intention to use the application.

Thus, the regression equation for predicting the behavioral intention to use the Markdown-to-Moodle application is:
 BI=.387*Error Tolerance+.225*Efficiency-2.389
  where BI is the behavioral intention to use the application.

# Discussion

Internet proficiency and experience are relevant to the utilization of the Markdown-to-Moodle Application. To maximize the Markdown-to-Moodle application, the teachers must be proficient in using different browsers, search engines, sources and hyperlinks. The teachers' familiarity with different browsers helps them to navigate the workspace easily. Moreover, the teachers can access the Markdown-to-Moodle through their mobile browsers. Teachers with experience in using search engines easily acquired images and used it in the test item construction. It is vital to embed the correct URL of the image; otherwise, it will not appear when running it on Moodle. Some teachers used their online data repository to save the images.

The significant improvement in constructing test items of the teachers in the Markdown-to-Moodle reveals the software's usability and how well the teachers remember the procedure and use the markdown symbols and format in item construction. This can be attributed to the reference guide which contains the instructions and rules located at the right portion of the application's interface. In case the teachers forget the symbols, format or procedure, they can easily read the reference guide.

The teachers strongly agree that the Markdown-to-Moodle Application is effective and engaging. Despite that Markdown-to-Moodle was newly introduced to teachers, the expected outputs such as the converted files in word, pdf and xml files were properly generated and imported to Moodle by the teachers. When the teachers run the imported test, Moodle processed the questions, displayed the images and generated the correct answer accurately. In addition, teachers perceived that the single workspace was engaging and could support their work in writing bulk or series of questionnaires for different subjects continuously. Teachers were pleased that the Markdown-to-Moodle could save the generated files such pdf, doc and xml from different sets of tests in one folder in the local drive. Furthermore, it generated a test questionnaire for students and a test questionnaire with answer key for teachers both in doc and pdf format. The generated documents can serve as a written report for teachers in compliance with the school's documentary requirements.

The teachers also agree that the Markdown-to-Moodle is efficient, error tolerant and easy to learn. Using Markdown-to-Moodle, teachers were able to quickly construct all the items in Set 2 without any supervision from the facilitators and easily import the converted xml questionnaire file to Moodle bank. The Markdown-to-Moodle assisted the teachers in learning the rules through the reference guide. Moreover, the Markdown-to-Moodle was able to recover the previous work done in the workspace. Even if the teacher accidentally closed the browser, the data can be retrieved when the application is reloaded. This error-tolerant capability of the Markdown-to-Moodle saves time and effort in reconstructing the sets of test items.

In the model, efficiency and error tolerance are the predictors that significantly affect the users' behavioral intention to use the Markdown-to-Moodle application. The continuity to do several tasks can be a key factor in the intention of the teacher to utilize the software in test construction. Markdown-to-Moodle eliminates the task of typing the questions and choices one by one in the Moodle text box. Thus, teachers can write test items continuously and quickly. This is beneficial for teachers who want to create set of tests for different subjects successively. Moreover, the teachers can just simply drag the xml file to the Moodle test bank which loads the test items instantly. The application's error-tolerant capability is also a contributing factor to the behavioral intention to use the application.

The Markdown-to-Moodle is a single page application. Hence, a user doesn't have to go to or load another page just to process another task which eliminates loading delays. Loading a page entails processing which consumes time depending on the internet connection speed and bandwidth (Rose, Lees, & Meuters, 2001). This loading delays interfere with the site's usability (Straub, Hoffman, Weber, & Steinfield, 2002) and waiting time is the most unpleasant application deficiency (Lightner & Bose, 1996). Thus, the teachers can accomplish their task without interferences and the time saved can be used to perform other tasks.

Moodle supports the GIFT format for constructing test items ("GIFT format," 2019). However, Markdown-to-Moodle can be utilized as a simpler and more readable alternative. Teachers must only utilize the asterisk (*), number sign (#), exclamation point (!) and dollar sign ($) symbols. Unlike in GIFT, several symbols must be used. In Markdown-to-Moodle, the indicator for correct answer are "correct" and "ans" which is a readable format while GIFT uses multiple symbols. Moreover, GIFT files must be correctly encoded in UTF-8 using a text editor while Markdown-to-Moodle serves as an editor and converter. Markdown-to-Moodle has the reference guide while GIFT format and instructions can be found at Moodle documentation site. GIFT matching type test format does not support feedback or percentage answer weights. However, Markdown-to-Moodle only supports multiple item response type. This can be a drawback or an advantage which may enforce teachers to write multiple choice questions which gives them the immediate evaluation response.

## Conclusion

The purpose of this study is to determine the performance of Markdown-to-Moodle application in easing the test construction process and explain the underlying factors of the behavioral intention to use the application.

In this study, the teacher internet proficiency and experience have a positive contribution to learning the Markdown-to-Moodle application. The teachers have positive acceptance of the Markdown-to-Moodle application. Teachers were also able to construct test items using Markdown format, convert the test items to *.doc, *.md and *.xml using the Markdown-to-Moodle application and import the xml file to the test bank. Also, they were able to run the test questions in Moodle correctly. Results reveal that the time of current task from the previous task in constructing, converting, and importing the test items is lesser, signifying learning gained from the intervention conducted.

Markdown-to-Moodle can generate a test questionnaire and a test questionnaire with answer key both in doc and pdf format and an xml files depending on the sets of test questionnaires. Thus, it addresses the teacher's effort on creating test questionnaires for different subjects as part of their written reports in compliance with the school's documentary requirements.

Meanwhile, the generated model in the analysis shows that efficiency and error tolerance significantly contribute to the teachers' behavioral intention to use Markdown-to-Moodle application. The continuity of workflow in test construction is a key factor both in the performance of the application and the teacher's effort to deliver and achieve the task quickly with minimal errors or problems.

While Markdown-to-Moodle can be an alternative tool in constructing numerous or bulk test items for different subjects addressing the tedious process of test construction in Moodle, it could only support a multiple-choice item response test. Hence, it is recommended to enhance the application's capability to support other test types like matching type, essay, and fill in the blanks. Likewise, since the number of respondents of the study from Indonesia was limited, future studies involving a larger sample size is encouraged.